\input{aipcheck}

\documentclass[
    ,final            
  ]
  {aipproc}

\layoutstyle{6x9}

\begin{document}

\title{Critical and resonance phenomena in neural networks}


%

\author{A. V. Goltsev}{
  address={Department of Physics $\&$ I3N, University of Aveiro, 3810-193 Aveiro, Portugal}
 ,altaddress={Ioffe Physico-Technical Institute, 194021, St. Petersburg, Russia} 
}

\author{M. A. Lopes}{
  address={Department of Physics $\&$ I3N, University of Aveiro, 3810-193 Aveiro, Portugal}
}

\author{K.-E. Lee}{
  address={Department of Physics $\&$ I3N, University of Aveiro, 3810-193 Aveiro, Portugal}
}
\author{J. F. F. Mendes}{
  address={Department of Physics $\&$ I3N, University of Aveiro, 3810-193 Aveiro, Portugal}
}

\begin{abstract}
Brain rhythms contribute to every aspect of brain function. Here, we study
critical and resonance phenomena that precede the emergence of brain rhythms. Using an analytical approach and simulations of a cortical circuit model of neural networks with stochastic neurons in the presence of noise, we show that spontaneous appearance of network oscillations occurs as a dynamical (non-equilibrium) phase transition at a critical point determined by the noise level, network structure, the balance between excitatory and inhibitory neurons, and other parameters. We find that the relaxation time of neural activity to a steady state, response to periodic stimuli at the frequency of the oscillations, amplitude of damped oscillations, and stochastic fluctuations of neural activity are dramatically increased when approaching the critical point of the transition.
\end{abstract}

\maketitle


\section{Introduction}

Brain rhythms contribute in every aspect of brain function from sensory and cognitive processing, and memory to motor control
\cite{Buzsaki_2006}. Origin and physiological functions of brain rhythms are a topic problem in neuroscience. Brain rhythms are also related to many unusual phenomena observed in the brain. 
Interactions between billions of neurons give rise to phase transitions, self-organization, and critical phenomena \cite{Chialvo_2006,Chialvo_2010}. Phase transitions were observed, for example, in human bimanual coordination \cite{Kelso_1984,Kelso_1985,Kelso_1986,Kelso_1987,Kelso_2010} and in living neural networks stimulated by electric fields \cite{Eckmann_2007}.
There are evidences that epileptic seizures, alpha and gamma oscillations, and the ultraslow oscillations of BOLD fMRI patterns emerge as a result of non-equilibrium phase transitions. Neural avalanches are one more example of critical collective phenomena observed in the brain \cite{Beggs_2003,Chialvo_2010}. 

Various resonance phenomena were also observed in the brain. Experimental investigations of CA1 neuronal networks from mammalian brain demonstrated that stochastic resonance can enhance effects of intrinsic 4-10 Hz hippocampal theta and 40 Hz gamma oscillations \citep{Gluckman_1996}. Recently, using a functional imaging technique, Sasaki \emph{et al.} \cite{Sasaki_2006} revealed that the majority of rat CA1 neurons act collectively like a band-pass filter. Damped oscillations and the Berger effect are also related to brain rhythms. The Berger effect manifests itself in activation of alpha waves on the electroencephalogram when the eyes are closed and diminution of alpha waves when they are opened \cite{Hari_1997}.


In the present paper, we study collective dynamics of neural networks composed by excitatory and inhibitory neurons in the presence of noise.
Based on exact analytical calculations and numerical simulations, we show that spontaneous emergence of network oscillations occurs as a dynamical (non-equilibrium) phase transition 
at a critical level of noise.
The  transition manifests itself in slowing down of the relaxation of a perturbed neural activity to a steady state, a strong enhancement of stochastic fluctuations of activities of neural populations and an increase of
the linear response function to afferent periodic stimuli at the frequency of neural oscillations. We show that near to the critical boundary, neural networks act as damped harmonic oscillators or  band-pass filters that pass frequencies within a certain range and attenuate frequencies outside that range.

\section{Cortical circuit model}

We use a cortical circuit model \cite{Goltsev_2010} composed of $N_e$ pyramidal cells (excitatory neurons) and $N_i$ interneurons (inhibitory  neurons) that form a sparsely connected network. The probability that there is a synaptic connection between two neurons is $c/N$ where $N=N_e+N_i$ is the total number of neurons and $c$ is the mean degree. This network has the structure of a directed
classical random graph (or Erd\H{o}s-R\'{e}nyi graph) with the Poisson degree distribution $P_{n}(c)=c^{n}e^{-c}/n!$ where $n$ is the number of presynaptic neurons.
Neurons receive sporadic inputs from a remote part of the cortex and synaptic noise.
Neurons fire with
a constant firing frequency $\nu$ that is the same for both excitatory and inhibitory neurons.
The total input $V_m$ to a neuron with index $m$, $m=1,2,\dots, N$,
is the sum of random spikes from
noise, excitatory  and inhibitory neurons,
\begin{equation}
V_m(t)=\sum_{n=1}^{N} k_n(t)a_{nm}J_{nm}+\xi(t),
\label{input}
\end{equation}
where $k_n(t)$ is the number of spikes that arrive from presynaptic neuron $n$ during the time interval $[t-\tau,t]$, $\tau$ is the integration time. Below we will consider the case $\tau \nu \leq 1$ when the number of spikes $k_n(t)$ is 1 or 0. If we assume that the emissions times of spikes of different neurons are uncorrelated, then the parameter $\tau \nu$ has a meaning of the probability that a postsynaptic neuron receives a spike from an active presynaptic neuron during time $\tau$.
Furthermore, $a_{nm}$ is the
adjacency matrix, i.e.,  $a_{nm}=1$ if there is a direct edge from neuron
$n$ to neuron $m$, otherwise $a_{nm}=0$. $J_{nm}$ is the efficacy
of the synapse connecting neuron $n$ with neuron $m$. $J_{nm}$ is positive if presynaptic neuron $n$ is excitatory and it is negative if the neuron is inhibitory. $\xi(t)$ is the number of random spikes from noise that neuron $m$
receives during the time interval $[t-\tau,t]$.  We use the Gaussian distribution for $\xi(t)$,
\begin{equation}
G(\xi)=A\exp \Big[-\frac{(\xi-\langle n\rangle)^2}{2\sigma^2}\Big],
\label{noise}
\end{equation}
where A is the normalization constant, $\sigma^2$ is the variance, $\langle n\rangle$ is the mean number of random spikes
determined by the mean rate $\omega_{rs}$, $\langle n\rangle = \omega_{rs}\tau$. Note that noise in our model is actually shot noise. According to Schottky's theorem, the intensity of this noise is proportional to $\langle n\rangle$.  

We consider stochastic neurons. Their
response on input is a stochastic process that occurs with a certain rate. 
Two rules determine dynamics of stochastic neurons \citep{Goltsev_2010}:
\begin{enumerate}
\item If the total input $V_m(t)$
at an inactive excitatory or inhibitory neuron $m$ at time $t$  is at
least a certain threshold $\Omega$ (i.e., $V_m(t)\geq \Omega$), then this
neuron is activated at a rate $\mu_{e}$ or $\mu_{i}$, respectively, and fires with a cyclic frequency $\nu$.
\item Active excitatory (inhibitory) neuron $m$ is inactivated at a
rate $\mu_{e}$ ($\mu_{i}$) if $V_m(t)< \Omega$.
\end{enumerate}
We assume that $1/\mu_{e}$ and $1/\mu_{i}$ are of the order of the first spike
latencies of excitatory and inhibitory neurons, respectively. We introduce the
ratio
\begin{equation}
\alpha \equiv \mu_{i}/\mu_{e}
\label{alpfa}
\end{equation}
that plays an important role in our model, as it will be shown below.
The advantage of this model with stochastic neurons is that it can be solved analytically.

In numerical simulations, we studied
sparsely connected networks of size $N=10^3-10^5$ and applied the following algorithm. We divided time $t$ into intervals of width $\Delta t=\tau$. At each time step, for each neuron, we calculated the input  Eq.~(\ref{input}), taking into account that each active presynaptic neuron contributes with a spike with probability $\tau \nu$. The number of random spikes from noise in this input is generated according to the Gaussian distribution, Eq.~(\ref{noise}). Then, with the probability $\tau \mu_a$, $a=e,i$, we updated the states of all neurons using the stochastic rules formulated above.
We used the following parameters:
the fraction of excitatory neurons is $g_e =N_e/N=75 \%$, the fraction of inhibitory neurons is $g_i=N_i/N=25 \%$,
the mean number of connections $c=1000$  (750 excitatory and 250 inhibitory connections), the threshold $\Omega=30$, and the variance of noise $\sigma^2 =10$. Following \cite{Amit_1997}, we chose $J_{ie}=J_{ii}\equiv J_i$, $J_{ee}=J_{ei}\equiv J_e$, and $J_{i}=-3J_{e}$.
These parameters agree with anatomical estimates for cortex. In cortex, the fraction $g_i$ of inhibitory neurons is between $0.15$ and $0.3$, the mean number of synaptic connections $c$
is about $7000$. The threshold $\Omega$ is between $15$ and $30$ in
neural networks {\it in vivo} \citep{Eckmann_2007} and about $30-400$ in the brain. The level of noise $\langle n\rangle$ was varied in the interval $0-150$ spikes per integration time $\tau$. We also assumed that, for simplicity, $\tau \nu =1$ and $\tau \mu_e =0.1$.

Dynamical behavior of the model
is described by the fractions $\rho_e(t)$ and $\rho_i(t)$ of active
excitatory and inhibitory neurons, respectively, at time $t$. We will call them
`activities' of the neural populations.
Using the rules of the stochastic dynamics formulated above and assuming that activities are changed slightly during the integration time $\tau$, in the infinite size limit $N\rightarrow\infty$, we find a rate equation \citep{Goltsev_2010},
\begin{equation}
\frac{d\rho_a (t)}{\mu_a dt}=f_{a}(t)(1{-}\rho_a(t))-\rho_{a}(t)+\Psi_a(\rho_{e}(t),\rho_{i}(t)).
\label{eq:10}
\end{equation}
for $a=e,i$. The function $\Psi_a(\rho_e,\rho_i)$ is the probability
that at time $t$ the input to a randomly chosen excitatory or inhibitory neuron
is at least the threshold $\Omega$. For the model under consideration $\Psi_i(\rho_e,\rho_i)=\Psi_e(\rho_e,\rho_i)\equiv
\Psi(\rho_e,\rho_i)$, where
\begin{equation}
\Psi(\rho_e,\rho_i)=\sum_{k=0}^{\infty}\sum_{l= 0}^{\infty}\sum_{\xi = -\infty}^{\infty}\Theta(J_{e}k {+}J_{i}l{+}\xi{-}\Omega) G(\xi) P_{k}(g_{e}\rho_{e} \widetilde{c})P_{l}(g_{i}\rho_{i} \widetilde{c}).
\label{eq:14}
\end{equation}
Here $\Theta(x)$ is the Heaviside step function, the parameter $\widetilde{c}$ is defined as
$\widetilde{c} \equiv c\nu\tau$, and
$P_k(g_e\rho_e \widetilde{c})$ and $P_l(g_i\rho_i \widetilde{c})$ are the probabilities that
a randomly chosen neuron receives $k$ spikes from active presynaptic excitatory and $l$ spikes from inhibitory neurons, respectively, during the time window $\tau$ at given activities $\rho_e$ and $\rho_i$. The functions $f_{e}(t)$ and $f_{i}(t)$ represent a rate of spontaneous activation of excitatory and inhibitory neurons, respectively, by stimulus, for example, an electric field.
The rate equation (\ref{eq:10}) is similar to the Wilson-Cowan equations \citep{Wilson_1972,Wilson_1973}, see also \citep{Goltsev_2010}. Equation (\ref{eq:10}) is asymptotically exact in the limit $N\rightarrow\infty$.

\begin{figure}
  \includegraphics[height=.3\textheight]{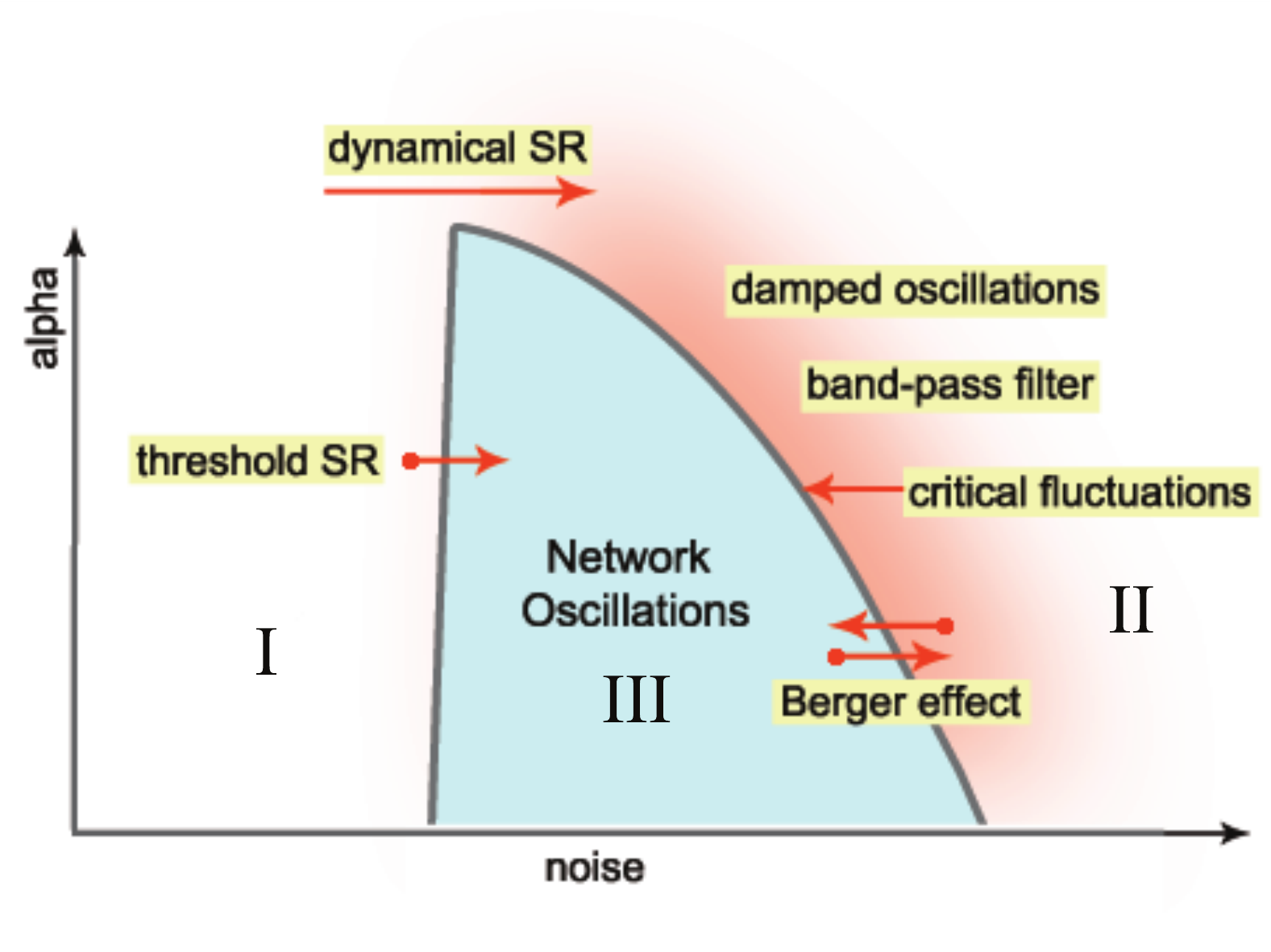}
  \caption{Schematic phase diagram of the cortical model and critical and resonance phenomena near the critical boundary of the non-equilibrium phase transition to sustained network oscillations. }
  \label{fig-overview}
\end{figure}

Steady states of the neural populations can be found from Eq.~(\ref{eq:10}), supposing $d\rho_{a}/dt = 0$ in the limit $t\rightarrow \infty$.
If $\rho_e(t)$ and $\rho_i(t)$ at time $t$ are close to steady state activities $\rho_e(\infty)$ and $\rho_i(\infty)$, then Eq.~(\ref{eq:10}) enables us to describe relaxation of $\rho_a(t)$ to the steady state. We introduce
\begin{equation}
\delta\rho_a(t)\equiv \rho_a(t)-\rho_a(\infty)
= Re (A_a e^{-\gamma t})
\label{relaxation}
\end{equation}
where $A_a$ is a complex amplitude.
Using the standard perturbation theory, we solve Eq.~(\ref{eq:10}) in the first order in $\delta\rho_a(t)$. We find 
\begin{equation}
\gamma_{\pm}=\frac{1}{2}(B_1+B_2)\pm \frac{1}{2}\Bigl[(B_1-B_2)^2+4\alpha D_{ei}D_{ie}\Bigr]^{1/2},
\label{eq-gamma}
\end{equation}
where we introduced parameters $B_1=1-D_{ee}$, $B_2=\alpha(1-D_{ii})$,
$D_{ab}=d\Psi_a(\rho_e,\rho_i)/d\rho_b$ for $a,b=e,i$.
respectively.
Derivatives $D_{ab}$ are
determined by the activities $\rho_e(\infty)$ and $\rho_i(\infty)$ from the non-linear equation Eq.~(\ref{eq:10}) when $d\rho_{a}/dt = 0$ \citep{Goltsev_2010}.
The real and imaginary parts of the complex rate $\gamma$ ($\gamma_r\equiv Re(\gamma_{-})$ and $\gamma_i\equiv Im(\gamma_{-})$) determine the relaxation rate and the angular frequency of damped oscillations, respectively. Notice that the period of the oscillations equals $2\pi/ \gamma_i$.

Analyzing behavior of $\gamma_r$ and
$\gamma_i$ in dependence on $\alpha$ and $\langle n\rangle$,
we obtain the phase diagram in Fig.~\ref{fig-overview}.
One can see that there are three regions. There is a region I (small noise level and/or large $\alpha$) where the relaxation of the neural activity to a steady state is exponential ($\gamma_r >0$ and $\gamma_i=0$). In region II,
the neural activity relaxes in a form of damped
oscillations ($\gamma_r >0$ and $\gamma_i \neq 0$).
In region III, network oscillations are sustained.
A similar phase diagram was found in \citep{Goltsev_2010}  for a simpler model.
If $\alpha$
is above a critical value $\alpha_t$, that corresponds to the $\alpha$-coordinate of the top point of the region III in Fig.~\ref{fig-overview}, then
with increasing the noise level $\langle n\rangle$, the activities $\rho_e$ and $\rho_i$ in the steady state undergo a first-order phase transition
at a critical noise level $n_c$.
A similar discontinuous transition was observed in living neural networks {\it in vitro} when living neural networks were stimulated by an electric field \cite{Eckmann_2007}. Neuronal avalanches are precursors of this phase transition. Activation (or inactivation) of one neuron can trigger avalanche  process of activation (or inactivation) of a cluster of neurons.
In cortex, neuronal avalanches have been observed experimentally \cite{Beggs_2003}, see the review  \cite{Chialvo_2010}.

If the parameter $\alpha <\alpha_t$,
sustained networks oscillations appear in a certain `optimal' range of the noise level $\langle n\rangle$ between two critical points.
Weak noise can not stimulate network oscillations. Too strong noise over-activates neural networks and only damped oscillations can occur.
The critical boundary of region with the sustained oscillations
is determined by the condition that the relaxation rate $\gamma_r$ is zero,
\begin{equation}
\gamma_r =Re(\gamma_{-})=0,
\label{gamma-r}
\end{equation}
where the complex frequency $\gamma_{-}$ is given by Eq.~(\ref{eq-gamma}).
For the parameters given above and $\tau=10$ ms,
frequencies of the oscillations lie in the range of brain waves (1-- 100 Hz).

\section{Linear response function and band-pass filter behavior}
\label{response function}

Now we study critical phenomena that precede the non-equilibrium phase transition from asynchronious dynamics to sustained oscillations. For this purpose we calculate the linear response of the neural network to a time-dependent stimulus $f_{e}(t)$ and $f_{i}(t)$ in Eq.~(\ref{eq:10}) for region I and II. Here we are not studying a response in region III that needs a special consideration.
A response of the neural population $a=e,i$ to a weak stimulus $f_{a}(t)$
is determined by the linear response function $\chi_{ab}(t-t')$,
\begin{equation}
\Delta\rho_a(t){\equiv} \rho_a(t) {-} \rho_a(\infty){=}\!\!\sum_{b=e,i} \!\!\int_{-\infty}^{t}\chi_{ab}(t{-}t') f_b(t')dt'.
\label{t-response}
\end{equation}
Solving Eq.~(\ref{eq:10})
in the linear-response regime, we find that in the regions I and II the neural network behaves as a damped oscillator driven by a force $F_e (t)$,
\begin{equation}
\frac{d^2 \Delta\rho_e(t)}{dt^2}+2\zeta \omega_0 \frac{d \Delta\rho_e(t)}{dt}+\omega_{0}^2 \Delta\rho_e(t)=F_{e}(t),
\label{d-oscillations}
\end{equation}
(see, for example, in \citep{Kubo_book}). Here we introduced the damping ratio $\zeta=\gamma_r /\omega_0$ and a frequency $\omega_0 =(\gamma_{r}^2 +\gamma_{i}^2)^{1/2}$.
In region I, the network  is critically damped because $\zeta =1$ and it is underdamped in region II, where $\zeta < 1$.
In the case $f_{e}(t)\neq 0$ and $f_{i}(t)=0$, the force $F_{e}(t)$ equals $F_{e}(t)=(1-\rho_{e}(\infty))(B_2 f_{e}(t)+ d f_{e}(t)/dt)$. The parameter $B_2$ was defined above.
Solving Eq.~(\ref{d-oscillations}) leads to a response function,
\begin{equation}
\chi_{ee}(t-t')=X_e e^{-\gamma_r (t-t')}\sin\Big[\gamma_{i} (t-t')+\Phi_e \Big].
\label{eq:y}
\end{equation}
where $X_e{=}(1{-}\rho_e(\infty))[1{+}(B_2 {-} \gamma_r)^{2}/\gamma_{i}^{2}]^{1/2}$
and $\Phi_e {=} \tan^{-1}[\gamma_i/(B_2 {-} \gamma_r)]$ (one finds a similar result for $X_i$ and $\Phi_i$ of inhibitory neurons). If $\gamma_r > 0$, then Eq.~(\ref{eq:y}) shows loss of memory in the neural network with increasing time interval $t{-}t'$. If $\gamma_r$ tends to zero, the memory becomes long-range.
The Fourier transform $\widetilde{\chi}_{ee}(\omega)$ of the linear response function is
\begin{equation}
\widetilde{\chi}_{ee}(\omega)=\frac{(1-\rho_e)(i\omega+B_2)}{\omega_{0}^2-\omega^2+2i\zeta  \omega_0 \omega}.
\label{Fourier}
\end{equation}
Equation~(\ref{Fourier}) shows that at $\zeta < 1$,
the neural network acts as a band-pass filter.
The spectral intensity  as a function of $\omega$ has a maximum at a resonance frequency $\omega_r \approx \omega_{0}\sqrt{1-2\zeta^2}$ at $\zeta < 1/\sqrt{2}$.
The maximum value $\|\widetilde{\chi}_{ee}(\omega_r)\|^2$
depends on the noise level $\langle n \rangle$.
When approaching the critical point, $\gamma_r \rightarrow 0$,
the value $\|\widetilde{\chi}_{ee}(\omega_r)\|^2$
diverges as $\widetilde{\chi}_{ee}(\omega_r) \propto 1/\gamma_{r}^2 \rightarrow\infty$, while the angular frequency of damped oscillations $\gamma_i$  tends to the frequency of stable network oscillations. This behavior signals that, in this regime, in the presence of noise, a neuronal network can amplify periodic signals. This amplification may be a mechanism of stochastic resonance observed in brain \cite{Moss_2004}.

The band pass filter behavior described by Eq.~(\ref{Fourier}) seems to be supported by
measurements of response of rat CA1 neurons to afferent stimulation \emph{in vitro} \citep{Sasaki_2006}. These measurements
revealed that the majority of rat CA1 neurons act collectively like a band-pass filter and fire synchronously in response to a limited range of presynaptic firing rates ($20-40$ Hz) that are in the range of gamma oscillations in the rat hippocampus \citep{Csicsvari_2003}.
One can also note that, a long time ago,
a number of characteristics of a band-pass filter behavior and a resonance response on sin wave trains already have been observed in EEG recordings of alpha activity \citep{Tweel_1964}.
Based on Eq.~(\ref{Fourier}), we suggest that band-pass filter behavior observed in \citet{Sasaki_2006} and \citet{Tweel_1964} is a manifestation of the critical phenomena near to the transition to neural network oscillations.

\section{Stochastic fluctuations of neuronal activity}
\label{fluctuations}

EEG measurements demonstrate that brain activity always contains a stochastic component. In this section we will show that stochastic fluctuations are enhanced when a neural network is close to the critical point of the non-equilibrium phase transition.
For characterizing stochastic fluctuations, we introduce the autocorrelation function
\begin{equation}
C_{ab}(t)=
\frac{1}{T}\int_{0}^{T}\delta\rho_a(t_{1})\delta\rho_b(t_{1}+t) dt_1,
\label{correl-1}
\end{equation}
where  $\delta \rho_a(t)=\rho_a(t)-\overline{\rho}_a$  describes fluctuations of activity $\rho_a(t)$ of population $a$, $a=e,i$, around the mean value $\overline{\rho}_a$   (see, for example, Ref.~\citep{Gardiner_2002}).
$C_{ab}(t)$ is a measure of correlations between values of $\delta \rho_a (t_1)$ and $\delta \rho_b(t_1 +t)$ at two different instants separated by a lag $t$ and averaged over an arbitrary large time window $T$. 
The Wiener-Khintchine theorem states that the power density spectrum of the fluctuations is the Fourier transform of the autocorrelation function.

For calculating the autocorrelation function, one uses the standard method \citep{Gardiner_2002,Thomas_1982}. In the deterministic equation (\ref{eq:10}), we assume that $f_a(t)$ is a stochastic force  that satisfies conditions $\langle f_a(t) \rangle=0$ and $\langle f_a(t) f_b(t')\rangle=f_{0}^2 \delta(t-t')\delta_{a,b}$.
If fluctuations are small, the autocorrelation function may be found in the linear response theory \citep{Gardiner_2002,Thomas_1982}. Assuming, for simplicity, $f_i(t)=0$, we obtain Eq.~(\ref{t-response}) that leads to
\begin{equation}
C_{ee}(t)=2\pi f_{0}^2\int_{-\infty}^{\infty}e^{i \omega t}\|\widetilde{\chi}_{ee}(\omega)\|^2 d\omega,
\label{correl-2}
\end{equation}
where the linear response function $\widetilde{\chi}_{ee}(\omega)$ is given by Eq.~(\ref{Fourier}).
In the region of damped oscillations, the autocorrelation function $C_{ee}(t)$ has a form
\begin{equation}
C_{ee}(t)=A_e e^{-\gamma_r |t|}\cos\Big(\gamma_{i} |t|+\Phi_e\Big).
\label{correl-4}
\end{equation}
The parameter $A_e$ and the phase  $\Psi_e$ behave as $A_e \propto 1 /\gamma_r$  and $\Phi_e \propto \gamma_r/\gamma_i$ at small $\gamma_r$. For inhibitory neurons we obtain a similar behavior. Thus, stochastic fluctuations
of activities of excitatory and inhibitory neural populations
are enhanced when approaching the critical point $\gamma_r=0$, Eq.~(\ref{gamma-r}), of the emergence of network oscillations (see Fig.~\ref{fig-overview}). However, the linear-response approximation is not valid when fluctuations become sufficiently large. This occurs near to the  non-equilibrium phase transition and non-perturbative methods are required for calculating $C_{ab}(t)$.

\section{Conclusion}
\label{conclusion}

In the present paper, using  a cortical model with stochastic neurons, we have showed that, in neuronal networks,  spontaneous appearance of sustained  network oscillations occurs as a non-equilibrium phase transition. The critical point is determined by the level of noise, structure of the neural network, the balance between excitatory and inhibitory neurons,  and other parameters. We have found critical and resonance phenomena that precede the transition. The important property of this transition is that, at the critical point, the relaxation time of the neuronal activity to a steady state becomes infinite in the infinite size limit.
An increase of the response of neural networks to periodic afferent stimulations and a strong enhancement of stochastic  fluctuations of activities of neural populations are also the critical phenomena that precede the transition.
Note, that these phenomena
are general properties of second-order phase transitions observed in physical, chemical and biological systems (see, for example, \citet{Stanley_book,Haken_1983,Kelso_2010}). These critical phenomena have been observed near the non-equilibrium phase transition in human hand movements  \cite{Kelso_1984,Kelso_1985,Kelso_1986,Kelso_1987,Kelso_2010}. The noise-induced nonequilibrium phase transition found in \cite{Toral1994} is one more example of a phase transition with similar critical phenomena.
Furthermore, we have demonstrated that near to the critical point, neuronal networks behave as damped harmonic oscillators or band-pass filters in agreement
with band-pass filter behavior observed \emph{in vitro} in networks
of CA1 neurons in mammalian brain \citep{Sasaki_2006}. We suggest that band-pass filter behavior is a manifestation of critical phenomena near to the transition to network oscillations.

We have also demonstrated that, in the cortical model, stochastic neural activity generated by a stochastic force is similar to spontaneous alpha activity
observed in EEG recordings of both a normal man
and human epileptic seizures of petit mal activity \citep{Babloyantz_1986}.


\begin{theacknowledgments}
  This work was partially supported by the following projects PTDC:
SAU-NEU/ 103904/2008, FIS/108476/2008, MAT/114515/2009, and PEst-C/CTM/LA0025/2011.
K.~E.~Lee was supported by FCT under grant SFRH/BPD/ 71883/2010,
M.~A.~L. was supported by FCT under Grant No. SFRH/BD/ 68743/2010.
\end{theacknowledgments}



\bibliographystyle{aipproc}   

\bibliography{Goltsev_proceedings}

\IfFileExists{\jobname.bbl}{}
 {\typeout{}
  \typeout{******************************************}
  \typeout{** Please run "bibtex \jobname" to optain}
  \typeout{** the bibliography and then re-run LaTeX}
  \typeout{** twice to fix the references!}
  \typeout{******************************************}
  \typeout{}
 }

\end{document}